# Substrate-effect on the magnetic microstructure of $La_{0.7}Sr_{0.3}MnO_3$ thin films studied by magnetic force microscopy

R. Desfeux, S. Bailleul and A. Da Costa

*Laboratoire de Physico-Chimie des Interfaces et Applications, Université d'Artois, Rue Jean Souvraz, SP 18, 62307 Lens Cedex, France*

W. Prellier and A.M. Haghiri-Gosnet

*Laboratoire CRISMAT, UMR CNRS 6508, 6 Boulevard du Maréchal Juin, 14050 Caen Cedex, France*

## ABSTRACT

Colossal magnetoresistive (CMR) $La_{0.7}Sr_{0.3}MnO_3$ (LSMO) thin films have been grown under tensile strains on (100)-$SrTiO_3$ substrates and compressive strains on (100)-$LaAlO_3$ and (110)-$NdGaO_3$ substrates by pulsed laser deposition. Using magnetic force microscopy (MFM), a "feather-like" magnetic pattern, characteristic of films with an in-plane magnetization, is observed for films deposited on both $SrTiO_3$ and $NdGaO_3$ while a "bubble" magnetic pattern, typical of films with an out-of plane magnetization, is recorded for $LaAlO_3$. We show that the shape of the magnetic pattern imaged by MFM is fully correlated to the easy direction of the magnetization in the film.

Electronic mail : desfeux@univ-artois.fr

The recent reports of spectacularly large magnetoresistance in the $R_{1-x}A_xMnO_3$ mixed valence manganite perovskite structures (R=rare earth, A=divalent alkali),[1-2] have renewed interest in studying these so-called "Colossal Magnetoresistance (CMR)" materials due to their potential applications in devices as field-sensor or magnetic reading heads. However, for a great number of these possible industrial applications, these materials have to be prepared in the form of thin films. For this reason numerous groups are working in order to get films with the best structural and physical properties.[3-4] However, the properties of the films are often different as compared to the bulk one. Thus, it is of a great interest to understand the origin of such differences. Recent results have shown that some properties, including specially structural deformation or anisotropic magnetic properties, were strongly dependent on the substrate.[5-6] Such results were explained in terms of stress (compressive or tensile) induced by lattice mismatch between the film and the substrate.

In this letter, we report on the characterization by x-rays diffraction (XRD), Atomic and Magnetic Force Microscopy (AFM and MFM) at Room Temperature (RT) of the surface morphology and of the magnetic microstructure of the ferromagnetic surface of $La_{0.7}Sr_{0.3}MnO_3$ (LSMO) thin films deposited on various substrates. We show that the substrate-induced stress leads to strong structural variations of the LSMO cell and that the different magnetic microstructures of the LSMO films can be fully correlated to the easy direction of magnetization M in the film.

LSMO thin films (1500Å) were grown by pulsed laser deposition. LSMO was chosen for this study since this compound appears to be one of the most attractive for device applications due to its high Curie Temperature (369 K).[7] Thus, this allows us undertaking MFM measurements at RT. High dense targets of LSMO were utilized for the deposition carried at 840°C under an oxygen pressure of to 0.3 mbar. After deposition, the films were slowly cooled to RT under 500 mbar of $O_2$.[3-4] The ferromagnetic state of the films at RT have been evidenced using a Quantum Design SQUID magnetometer. The substrates are (100)-SrTiO$_3$ (STO, cubic with a=3.905Å), (100)-LaAlO$_3$ (LAO, pseudocubic with a=3.789Å) and (110)-NdGaO$_3$ (NGO, orthorhombic with a=5.426Å, b=5.502Å and c=7.706Å). To eliminate extrinsic factors during the growth, the different samples were deposited in the same run.

Structural characterization was performed in the θ-2θ scan mode using a Rigaku Miniflex + diffractometer (CuKα1, λ=1.5406Å). AFM and MFM measurements were carried out in air at RT by using a Park Autoprobe CP Scanning Force Microscope. The AFM images were done in contact mode with Si ultralever tips. MFM imaging was performed in non-contact mode using the ac method of detection which is responsive to force gradients. Si microlevers tips with an overcoat of cobalt have been used. Before imaging, the tip is magnetized in a DC magnetic field in order to align the moments of the tip perpendicular to the sample, i.e. along the growth direction of the thin film. For this reason the MFM contrast will be higher on films presenting an out-of-plane easy axis. Moreover, it will be more difficult to image the magnetic pattern of films with an in-plane easy axis and without any out-plane magnetic component.

Considering that the LSMO materials deposited in form of thin films do not exhibit anymore a rhombohedral cell as the bulk one's but a tetragonal one with a I-type lattice leading to films with the $\vec{c}$-axis perpendicular to the plane of the substrate as reported by Haghiri-Gosnet et al,[4] we can ensured, through the XRD patterns that [001]$_{LSMO}$//[001]$_{STO\ and\ LAO}$ and [001]$_{LSMO}$//[$\bar{1}$10]$_{NGO}$. Note that in all the films, no other impurity phase is detectable. The $\vec{c}$-axis lattice constant calculation of the film shows considerable deviations from the bulk LSMO value (3.889 Å). On STO, the $\vec{c}$-axis lattice constant value is shortened (c = 3.851Å) while on LAO and NGO, it is elongated (3.919Å and 3.902Å respectively). From the calculation of the lattice mismatch δ between the substrate and the LSMO bulk material (δ =

+0.41% for STO, -2.57% for LAO and –0.78% for NGO, we show that the $\bar{c}$-axis lattice constant values are in good agreement with the Poisson's relation i.e. when the *a-* and *b-* lattice constants values in the plane of the substrate are bigger than those of the LSMO bulk (case of STO), the films grown under tensile strains whereas when *a-* and *b-* are smaller (case of LAO and NGO), the films grown under compressive strains. Such results, in good agreement with Tsui *et al* on LSMO films grown on various substrates,[8] are not surprising since in fact, these manganites thin films are very sensitive to external perturbations, strain being one of them.[9]

From the AFM study performed on the different samples, we can see that the surface morphology is similar for all the films with the existence of a granular-like surface with rounded grains (Fig. 1). This similarity in the growth morphology on the different substrates, showing an island-like growth mode, is surprising. This means that during the ablation process, the LSMO accommodates the strains, leading to strong distortion of the perovskite cell. As a result, the film of oxide will grow in the same way since its cell is highly distorted. However, some differences exist concerning the mean grains diameter, the mean surface roughness ($R_{ms}$) and the maximum peak to valley roughness ($R_{p-v}$) values depending on the choice of the substrate. On LAO and NGO, we remark that the $R_{ms}$ and $R_{p-v}$ values are about the same (40 Å and 280 Å respectively) but are lower that those obtained for the films on STO which are about the double ($R_{ms}$=80 Å and $R_{p-v}$=490 Å). Concerning the mean grains diameter values, we found that they are in the same order as the $R_{ms}$ and $R_{p-v}$ values. Indeed, the mean grains diameter of films grown on STO is around 750 Å i.e . larger than this obtained on NGO (700 Å) and LAO (600 Å). Kwon *et al*.[6] have attributed theses results to the difference in the atomic mobility during growth between the different substrates. Indeed, the surface mobility usually results in improvements in film-surface morphology because of enhancement nucleation and annealing defects during growth.[10] Thus in our films, we conclude the mobility of the species is higher on STO substrates than on NGO and LAO substrates leading to a higher roughness on STO, in spite of the mismatch between the film and the substrate. This also indicates that in LSMO films, the mobility is the major factor for the roughness before the mismatch. Next, the samples which have never come into contact with a magnetic field have been studied by MFM. The MFM images (Fig. 2) show that there is no direct correspondence with the granular-like surface observed on the AFM images (Fig. 1). Such an observation confirms that there is no topographic information on the MFM images and that the observed contrasts are only due to the sample-tip magnetic interaction. On the Figs. 2(a) and 2(c), some artefacts appear on the form of horizontal lines. Such artefacts can be explained by the strong difficulty to stabilize the MFM signal or/and by a switching of magnetic domains due to the interactions with the tip during the scan confirming the small coercitive field of the film ($H_c \approx 5$ Gauss[4,6,8]). The MFM images are totally different on each substrate. Thus, on both STO and NGO, we can see an uniform color surface image, without significant contrasts [see Figs 2(a) and 2(c)]. Such a magnetic pattern, described as a "feather-like" pattern by Kwon et al,[6] is typical of films with an in-plane magnetization.[11] For LAO substrates, we can see highly circular contrasted black and white domains which have the shape of "bubble". Such patterns are characteristics of films with an out-of-plane magnetization. The contrast comes from domains with a magnetization up for the white domains and down for the black domains indicating a large perpendicular anisotropy. The mean magnetic circular domains diameter is about 0.3 µm. Such a stable and spontaneous appearance of magnetic bubble domains without an external magnetic field has been already seen in a $La_{1.4}Sr_{1.6}Mn_2O_7$ layered ferromagnetic manganite crystal.[12] This feature has been shown as extremely interesting in order to develop devices based on magnetic bubble memory. However to our knowledge, this phenomenon was only observed at low temperature

(close to 70 K). So, the results obtained on LSMO films deposited on LAO are remarkable in the sense that this phenomenon can be observed at RT without external magnetic field.

To explain the role of the substrate on the shape of the magnetic domains observed by MFM at the surface of the film, we have considered the values of the lattice mismatch δ and the easy direction of magnetization M in the film. It is shown that a positive value of δ (for STO) led to a magnetic pattern with M in the plane (case of STO).[4-6] On the contrary, when δ was negative (for NGO and LAO), two magnetic patterns were observed i.e. a one with M out-of-plane (for LAO) and one other with M in the plane (for NGO). These results show that the magnetic pattern shape can not be correlated to the negative or positive value of δ since a similar « feather-like » magnetic pattern is obtained for films under tensile (case of STO) and compressive strain (case of NGO). A more relevant parameter to predict the shape of the magnetic pattern is the easy direction of magnetization M of the films. LSMO films deposited on STO and NGO exhibit an easy-plane magnetic anisotropy[4,6,8], while on LAO films present an out-plane anisotropy[4,6]. As a conclusion, the magnetic pattern in LSMO films observed by MFM can be explained by taking into account the direction of the easy axis direction of magnetization of the film : an in-plane magnetization will lead to a "feather-like" magnetic pattern while an out-of-plane magnetization will lead to a "bubble" magnetic pattern. However, interestingly is the magnetic domains shape when M is out-of-plane in the case of LAO. Indeed, the "bubble" magnetic domain shape obtained on LAO is quite different compared to the "maze-like" 's one recorded by Kwon *et al.*[6] on such a type of sample. Although these two patterns are characteristics of films with M out-of-plane [13], the explanation of such a magnetic domain shape is not so easy except if we take into consideration the nature and the structural properties of LAO substrates.[14] Due to the phase transition between the cubic prototype high temperature phase above $T_C$ = 544°C and rhombohedral symmetry below $T_C$, twinning appear in the {001} and {110} planes of the pseudo-cubic lattice when temperature is decreasing below $T_C$ during the cooling after growth. At RT, the crystal is cut and polished near a rational crystallographic plane i.e. $(001)_{cub}$. When the substrate is heated above $T_C$ for the growth of the film, the twinning disappear. However, "footprints" of the previous twin-domain pattern will remain as a surface relief. Also, during the growth, thermal strains or other defects will be redistributed in the substrate due to the growth temperature used (around 800°C). Twin walls will occur in other positions and orientations when the crystal will be again cooled to RT. So the crystal will be an overlay of new twin walls and footprints of previous domains. So, we suggest that the growth conditions play an important role on the structural properties of the LAO substrate leading to the observation of two different magnetic microstructures ("maze-like" and "bubble") at the surface of the film.

In conclusion, we have demonstrated that lattice distortions of CMR LSMO thin films were obtained on various substrates. We have shown that the magnetic pattern recorded by MFM traduces the magnetic anisotropy of the film, i.e. M in or out-of plane. Moreover, the observation of spontaneous "bubble magnetic" domains, at room temperature and without external field, in the case of LAO substrates open a potential application of these CMR materials for high density recording and magnetic bubble memories.

FIGURE CAPTIONS

FIG. 1 : 3D AFM images showing the granular surface of LSMO thin films deposited on STO (a), LAO (b) and NGO (c) substrates. The scan size is 1 µm x 1µm.

FIG. 2 : MFM images of LSMO thin films deposited on STO (a), LAO (b) and NGO (c) substrates recorded at RT without external field. The scan size is 5 µm x 5µm.

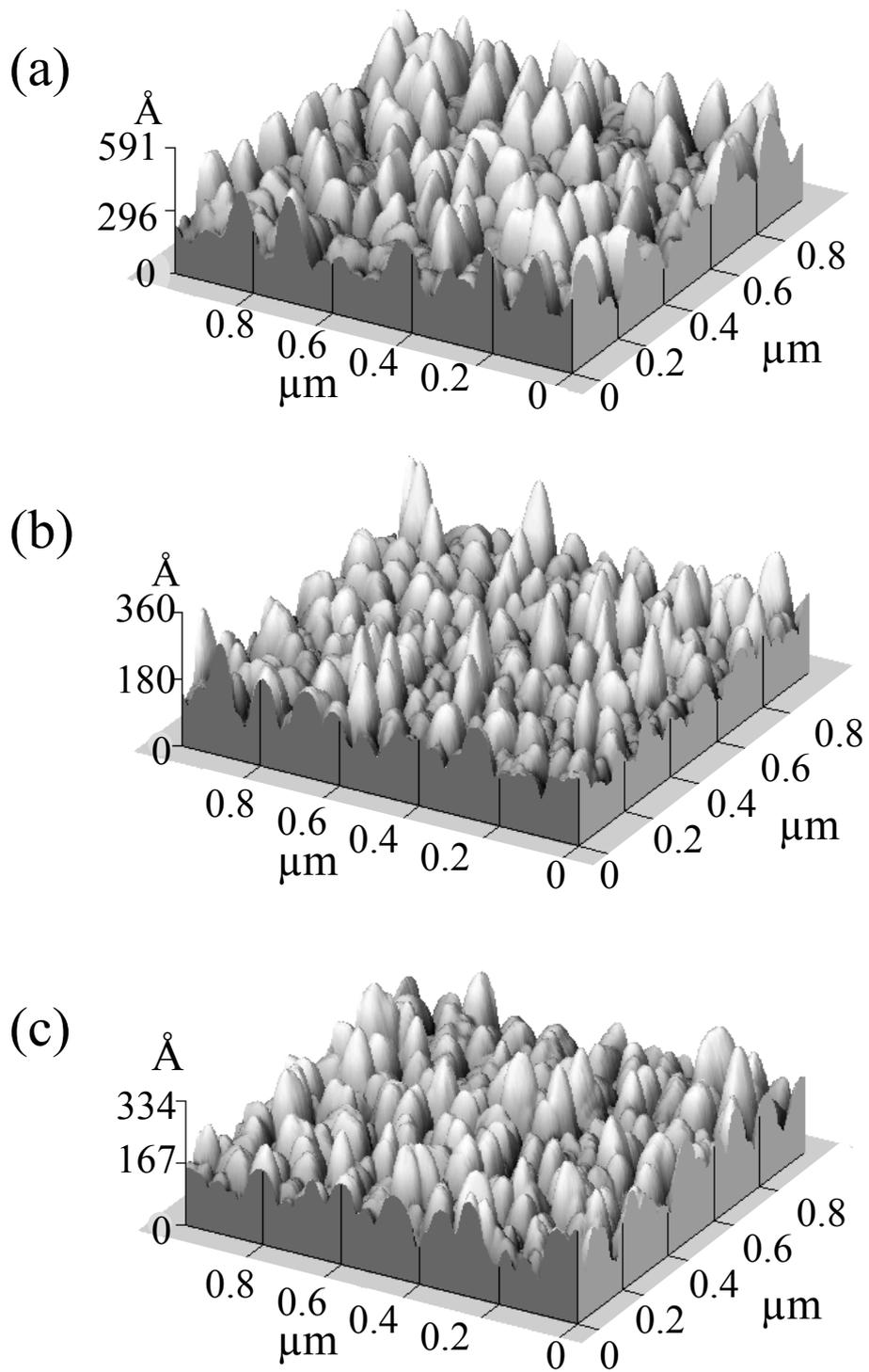

Figure1

(a)
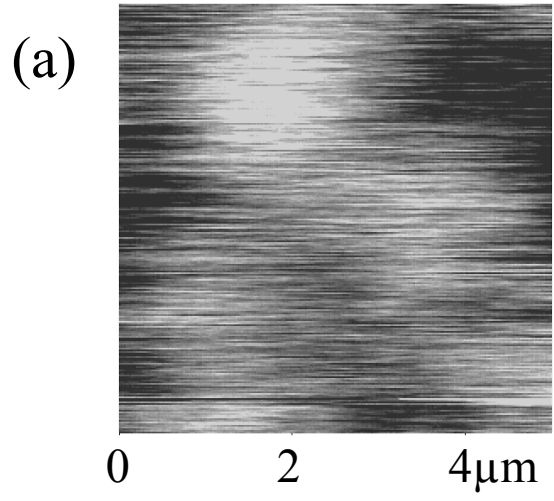
0    2    4μm

(b)
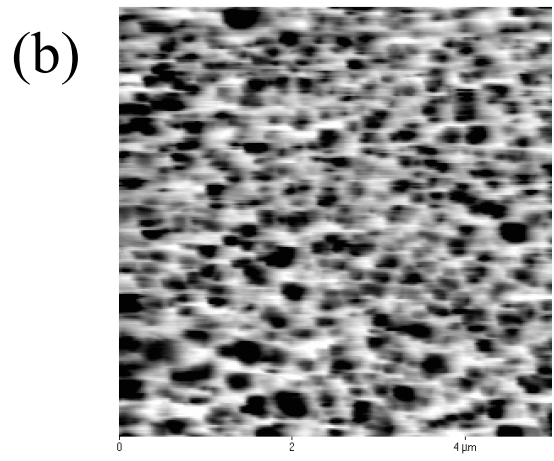

Figure 2